\documentclass[universe,article,accept,pdftex,moreauthors]{Definitions/mdpi} 
\firstpage{1} 
\pubvolume{1}
\issuenum{1}
\articlenumber{0}
\pubyear{2025}
\copyrightyear{2025}
\datereceived{22 January 2025 } 
\daterevised{26 March 2025 } 
\dateaccepted{4 April 2025 } 
\datepublished{ } 
\hreflink{https://doi.org/} 

\Title{Scanning the Universe for Large-Scale Structures \\ using Gamma-Ray Bursts} 
\TitleCitation{Scanning the Universe using Gamma-Ray Bursts} 

\Author{{Istvan Horvath} 
 $^{1}$\orcidA{},
Zsolt Bagoly $^{1,2}$\orcidB{},
Lajos G. Balazs $^{3,4}$\orcidC{},
{Jon Hakkila} 
 $^{5}$\orcidD{},
Bendeguz Koncz $^{3,6}$\orcidF{},
Istvan I. Racz $^{1}$\orcidH{},
Peter Veres $^{7,8}$\orcidG{} 
and 
Sandor Pinter $^{1,}$*\orcidE{}
}

\AuthorNames{Istvan Horvath, Zsolt Bagoly, Lajos G. Balazs, Jon Hakkila, Bendeguz Koncz, Istvan I. Racz, Peter Veres, and 
Sandor Pinter}

\isAPAStyle{%
       \AuthorCitation{Lastname, F., Lastname, F., \& Lastname, F.}
         }{%
        \isChicagoStyle{%
        \AuthorCitation{Lastname, Firstname, Firstname Lastname, and Firstname Lastname.}
        }{
        \AuthorCitation{Horvath, I.; Bagoly, Z.; Balazs, L.G.; Hakkila, J.; Koncz, B.; Racz, I.I.; Veres, P.; Pinter, S.}
        }
}

\address{$^{1}$ \quad {Department of Natural Science, }University 
 of Public Service, 
{H-1441} Budapest, 
Hungary; {horvath.istvan@uni-nke.hu (I.H.); zsolt.bagoly@elte.hu (Z.B.); racz.istvan@uni-nke.hu (I.I.R.)} 
\\
$^{2}$ \quad Department of Physics of Complex Systems, E\"otv\"os University, {H-1053} Budapest, Hungary;
\\
$^{3}$ \quad Department of Astronomy, E\"otv\"os University, {H-1053} Budapest, Hungary; {balazs@konkoly.hu (L.G.B.); kbendeg398@mailbox.unideb.hu (B.K.)}\\
$^{4}$ \quad Konkoly Observatory, Research Centre for Astronomy and Earth Sciences, {H-1121} Budapest, Hungary\\
$^{5}$ \quad Department of Physics and Astronomy, University of Alabama in Huntsville, Huntsville, AL 35899, USA; {jh0271@uah.edu}\\
$^{6}$ \quad Faculty of Science and Technology, University of Debrecen, {H-4002} Debrecen, Hungary\\
$^{7}$ \quad Department of Space Science, University of Alabama in Huntsville, Huntsville, AL 35899, USA; {peter.veres@uah.edu}\\
$^{8}$ \quad Center for Space Plasma and Aeronomic Research, University of Alabama in Huntsville, Huntsville, AL {35805}, USA\\}

\corres{Correspondence: sandor.pinter@uni-nke.hu}

\abstract{In the past few decades, large universal structures have been found that challenge the homogeneity and isotropy expected in standard cosmological models. The largest of these, identified as the Hercules--Corona Borealis Great Wall, was found in 2014 in the northern galactic hemisphere in the redshift range of $1.6 \le z \le 2.1$. Subsequent studies used an increasing gamma-ray burst database to show that the cluster was unlikely to have been caused by statistical sampling uncertainties. This study re-examines burst clustering in the northern galactic hemisphere using a recently developed methodology. Evidence is provided that the Hercules--Corona Borealis Great Wall cluster is larger than previously thought, with members potentially spanning the redshift range of $0.33 \le z \le 2.43$. \textls[-5]{The extension of this cluster's size does not appear to have been due to statistical variations or sampling biases.} }


\begin{document}

\section{Introduction}

Gamma-ray bursts (GRBs) are among the most luminous events observed in the universe. For decades, their immense energy outputs have captivated scientists, and they remain enigmatic given the many unanswered questions regarding their origins \citep{2015AdAst2015E..22P,zhang_2018,2022GalaxBosPeer}. 
It is generally accepted that long-duration GRBs result from massive star collapse \citep{woo93,wb06} and that short-duration GRBs originate from compact stellar mergers in binary systems \citep{berger14}.  Both progenitor types can produce ultra-relativistic beamed jets capable of converting kinetic energy into high-energy photons on very short timescales.

GRBs have large luminosities that allow them to be seen at great distances and long lookback times. Using current observational technology, GRBs can be detected at z $\approx$ 7 or even higher redshifts \citep{2015A&A...581A..86M,Salvaterra_2009,2006Natur.443..186I}. If the universe exhibits large inhomogeneities, then GRBs can conceivably be used to map large-scale universal structures \citep{mp06,2019MNRAS.490.4481And,2020MNRAS.498.2544H,2023ApJ...952....3T,2023MNRAS.521.3909B,2024MNRAS.533..743W,Dainotti24MNRAS533}. Early studies of GRBs found the sky distribution isotropic \citep{Briggs96,tarno16MNRAS,2019ApJTarn,2019MNRAS.486.3027Ripa,2019MNRAS.490.4481And,2022AATarn} (although some subsamples showed slight anisotropy \citep{bal99,mesz00,mgc03,vbh08,R_pa_2017,2023MNRAS.524.3252H,2023CQGra..40i4001K}). 
However, these studies could not easily probe universal homogeneity because early instruments were unable to measure GRB distances. This issue was resolved with the discovery that redshifts could be obtained from observations of GRB \mbox{afterglows \citep{1997Natur.387..783C}.}
 
According to \cite{2010MNRAS.405.2009Y}, the theoretical scale on which the universe transitions from heterogeneous to homogeneous is $260 h^{-1}=356$ Mpc (where $h$ is the dimensionless Hubble parameter having a value of $h=0.73$). Any structures larger than this would not have had time to form without violating the cosmological principle. Curiously, surveys of luminous astronomical objects such as gamma-ray bursts have increasingly identified structures exceeding this theoretical limit. The structures that have been found resulted from the clustering of galaxies (the Sloan Great Wall \citep{Gott05}, the South Pole Wall \citep{2020ApJ...897..133P}, and the King Ghidorah Supercluster \citep{2023MNRAS.519L..45S}), quasars (the Clowes--Campusano Large Quasar Group \citep{1991MNRAS.249..218C}, the U1.11 Large Quasar Group \citep{2012MNRAS.419..556C}, the Huge Large Quasar Group \citep{clo12}, the Giant Quasar Arc \citep{2022MNRASLopez}, and the Big Ring \citep{2024JCAP...07..055L}), and GRBs (the Giant GRB Ring \citep{BalazsRing2015,BalazsTus2018} and the Hercules--Corona Borealis Great Wall \citep{hhb14,hbht15,2020MNRAS.498.2544H}). 
These structures challenge the cosmological principles of homogeneity and isotropy \citep{Li2015}, making their study crucial for understanding cosmic \mbox{structures \citep{2021A&A...653A.123M,2021MNRAS.507.1361M,2021A&A...649A.151M,2022SerAJ.204...29F}. }

The Hercules--Corona Borealis Great Wall (HerCrbGW) is the largest of these structures. However, as GRBs are transient events, only part of this structure can be seen at any given time, and its true size and shape remain unclear. Bursts observed over integrated time spans provide larger samples that can be used to identify the structure relative to the \mbox{universal average. }

The HerCrbGW was found in 2014 \citep{hhb14,hbht15} in a sample of 283 GRBs with known redshift. This GRB cluster occupies a large portion of the northern galactic hemisphere in the redshift range of $1.6 \le z \le 2.1$. A larger sample of 487 GRBs was subsequently studied using the point radius method \citet{2020MNRAS.498.2544H}, verifying that the cluster was unlikely to have been caused by statistical fluctuations in the GRB detection rate and extending its radial range to $1.6 \le z \le 2.3$. 
In 2024, a re-application of the point radius method to 542 GRBs with known redshift \cite{horv24MNRAS} recovered several significantly large clusters, one of which was the HerCrbGW group, and another was a further extension of it at $0.9 \le z \le 2.1$. 
Although \cite{horv24MNRAS} scanned the sky for $z$ slices as large as 119 GRBs, no attempt was made to determine whether or not the northern galactic hemisphere HerCrbGW cluster extended to a larger scale. 
Since 262 of the aforementioned 542 GRBs are in the northern galactic hemisphere, in this paper we describe our search for evidence of clusters with z slices as large as \mbox{242 GRBs.}

\section{Data and Methods}

This study utilized a database of gamma-ray bursts with measured positions on the celestial sphere, optical afterglows, and spectroscopic redshifts. A majority of these were detected by NASA's Swift and Fermi experiments. This was the same dataset used by \cite{2022Univ....8..221H,universe8070342,horv24MNRAS}. 
Most of the redshifts were primarily obtained from the Gamma-Ray Burst Online Index (GRBOX) database  \endnote{{published} 
by the Caltech Astronomy Department:  \url{http://sites.astro.caltech.edu/grbox/grbox.php}  {(accessed on 15 January 2024)}}, 
which was discontinued \mbox{in 2018.}

The data from \cite{2020MNRAS.498.2544H} were supplemented with new observations made up until \mbox{31 August 2022,} resulting in a total of 542 GRBs with accurately measured redshifts and known angular locations. 
The GRBs' redshift detection completeness was about\mbox{ 41 percent \citep{2021MNRAS.508...52L};} taking only the spectroscopic redshifts into account, the completeness was reduced to 30 percent.
The datasets were updated from different sources in the public domain, mainly from the Gamma-ray Coordination Network (GCN) and the publicly available dataset compiled by Jochen Greiner \endnote{\url{https://www.mpe.mpg.de/~jcg/grbgen.html} {(accessed on 15 January 2024)}} that provides extensive information on nearly all GRBs observed by \mbox{any instrument. }

The data exclusively consisted of spectroscopic redshifts, disregarding photometric redshifts and redshift estimates based on Ly-alpha limits, as these tend to introduce significant radial distance uncertainties generally exceeding several hundred Mpcs.

The dataset used was the most extensive one available for global studies, containing refined GRB distances obtained from spectroscopic redshifts. 
The possibility of significant statistical improvements in the near future 
will be low, as the number of new redshift detections rapidly decreased over the past few years \citep{universe8070342}. 


{Aside from the fact that GRBs are not standard candles in the gamma band} \cite{2015MNRAS.453..128L,2019ApJ...887...13F,2024A&A...689A.165L}, {there are much more complex observational effects at stake.} 
The probability of the successful detection of redshifts clearly depends on the very existence of an optical afterglow, the geometry, the type of the progenitor, and the ground observational factors (telescope availability, visibility, timing, etc.). According to \citet{2013arXiv1309.3988J}, the fraction of Swift-detected GRBs at a high redshift was below $10\%\, (5\%)$ for $z > 6\, (z > 7)$. 
Beside the spatial distribution of the GRBs, all the above factors shape the observed redshift distribution. We have shown that the ground factors are not trackable, and the data cannot be corrected for these effects \citep{universe8070342}. All this means that the database would have been shaped by observational effects, and only statistical methods dealing with these variables were applicable for our study.

We would also like to mention that our previous work has shown \cite{2023CoSka..53d.115R} that the T90 duration of GRBs is strongly correlated with the redshift, but not with the peak flux. The fluence also correlates with the T90 and redshift. The relationship is not a simple time dilation, and different brightness distributions result in different sample volumes. Galactic foreground extinction affects the sky distribution, with some groups showing an increased density.

\subsection{Redshift and Angular Distributions}

\begin{figure}[H]
    \includegraphics[width=0.97\columnwidth]{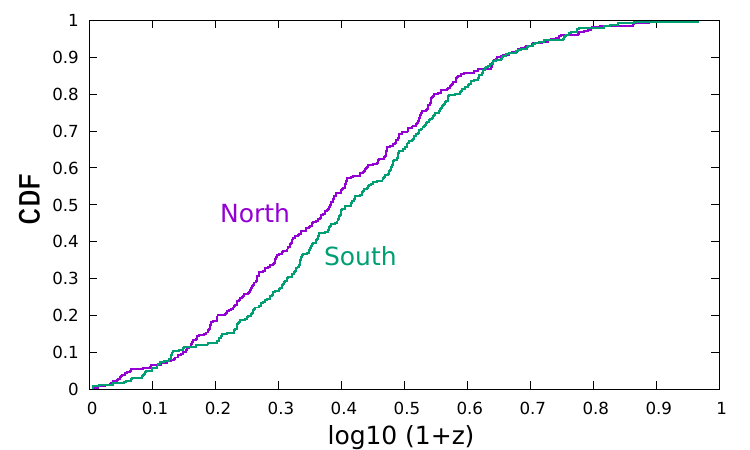}
    \caption{The cumulative distribution function (CDF) of $\log_{10}{(1+z)}$ for the GRBs in this study. Purple indicates the 262 GRBs found in the northern galactic sky, while green indicates the 280 GRBs found in the southern galactic sky. The difference between the two CDFs is insignificant (see text).}
    \label{fig:cumdist}
\end{figure}

The radial coordinates of the 542 GRBs were obtained from their observed redshifts, which relate to the radial distance via equations governing the cosmological expansion of the universe. The cumulative distribution of these redshifts is demonstrated in Figure~\ref{fig:cumdist}. Of these, 262 were found in the northern galactic hemisphere, while the remaining 280 were found in the southern galactic hemisphere. A two-sample Kolmogorov--Smirnov test of these redshift distributions showed no significant difference between the distributions in the two hemispheres ($D=0.098$, $p=0.151$).

\begin{figure}[H]
    \includegraphics[width=0.97\columnwidth]{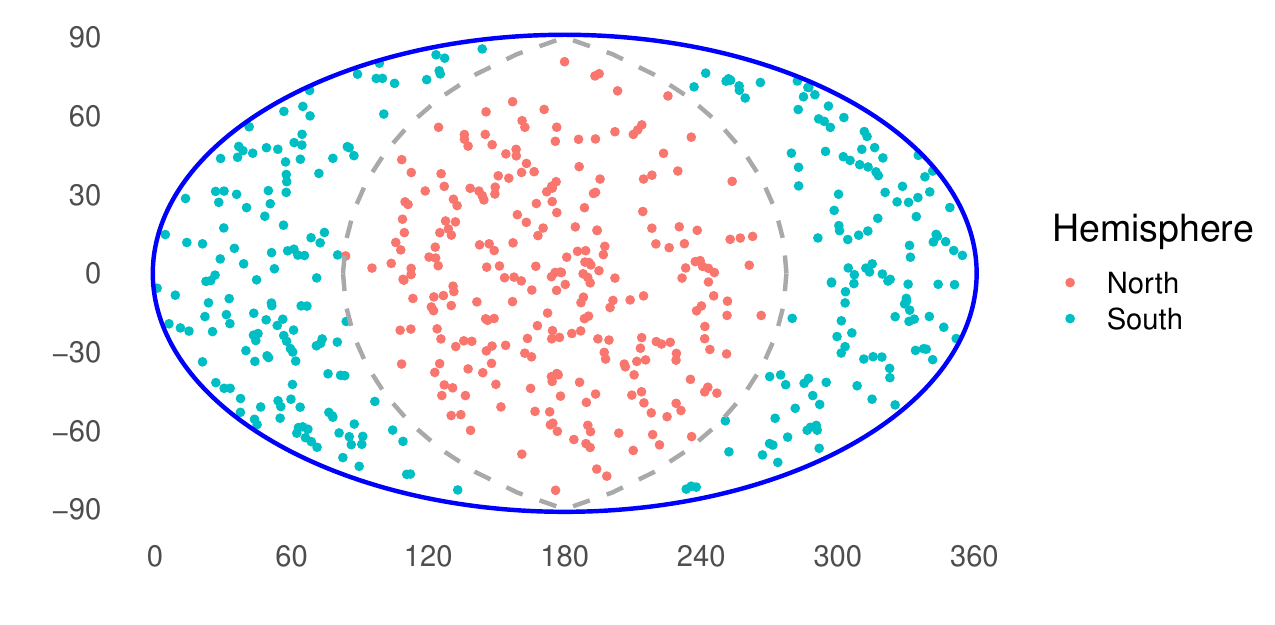}
    \includegraphics[width=0.97\columnwidth]{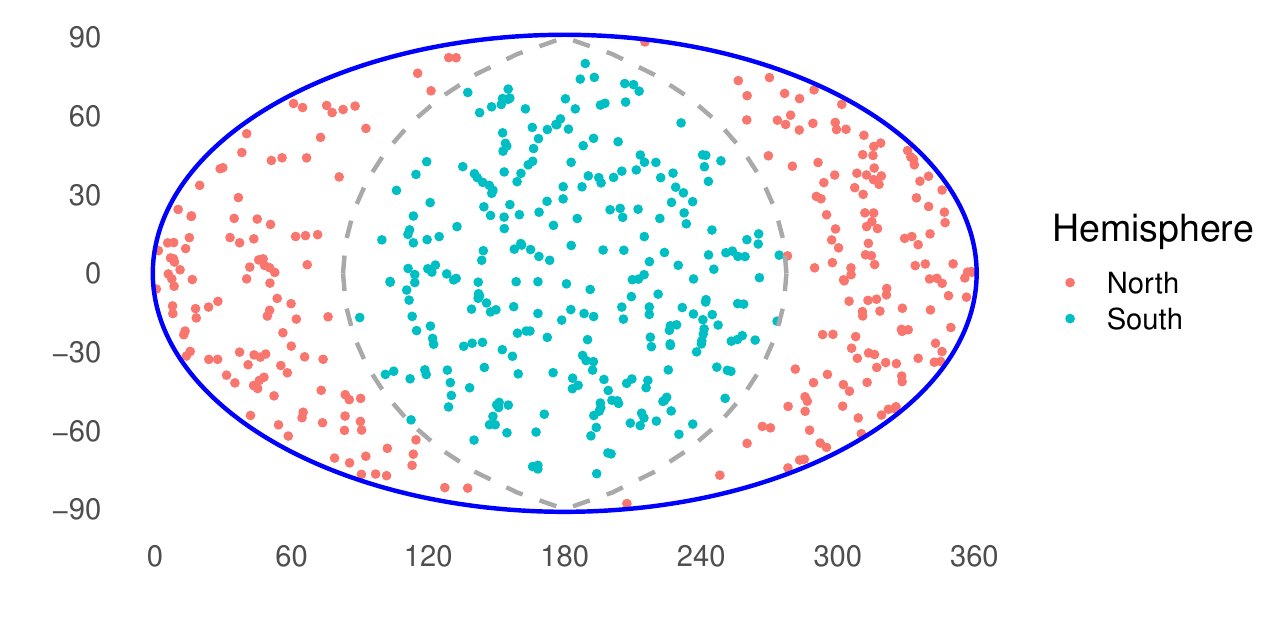}
    \caption{The sky distribution of the 542-GRB sample viewed from the galactic poles. The top/bottom figure is centered on the northern/southern galactic pole. The gray dashed line represents the location of the galactic equator.}
    \label{fig:galacticnorthern}
\end{figure}

The angular distribution of the 542 GRBs on the plane of the sky is shown in Figure~\ref{fig:galacticnorthern}. Due to galactic extinction, spectroscopic redshifts can only be measured for a small number of GRBs near the galactic equator, which results in noticeable GRB depletion near the galactic equator (identified by the dashed line in Figure~\ref{fig:galacticnorthern}). This area of reduced observational probability makes estimating the GRB density distribution across the plane of the sky much less certain. Reference \cite{horv24MNRAS} circumvented this uncertainty by analyzing the two galactic hemisphere distributions independently. Because the HerCrBGW is confined to GRBs in the northern galactic hemisphere, in this work we analyzed only the northern hemisphere GRBs' redshift and angular distributions.

\vspace{-12pt}

\subsection{Analysis Method}
\label{sec:method}

The bootstrap point radius test can be used to analyze the distribution of points on a surface. It identifies regions in which point densities deviate significantly from the mean.
This technique, which has the ability to take known GRB sampling biases into consideration, has been used to search for clusters in the GRB spatial distribution \cite{hhb14,hbht15}. The approach is as follows:
Starting at location $k$, a slice of $n$ successive GRBs is selected in the radial ($z$ or co-moving distance) distribution. Each location, $k$, is used as the center of a circle (spherical cap) specified by a spherical surface, $A$.
For all spherical cap center positions on the sky, and for all radial starting positions, the maximum number of GRBs within the circle, 
$K$, is found. The parameter $K$ is dependent only on the selected $A$ cap area and $n$ radial slice size parameters. 

Note that this test does not use the direct distance measurements of the events ({e.g.,} 
 co-moving distances). The radial slice approach used here depends only on the relative order of GRB distances such that ordering the points by the redshift is sufficient. For all physical distances, we used co-moving distances. 

This method is one of the best for 
adaptively changing the sensitivity according to the density. Where the density is high, the probability of detecting a density anomaly will be higher, and it will be lower in low-density areas due to the noise level.

For the almost-uniform spherical cap center positions, a HEALPix partition \citep{2005ApJ...622..759G} was selected.
The resolution of HEALPix pixelization is given by its $N_{side}$ parameter. We used $N_{side}=6$ for pixelization to obtain $12 \times ( 2^6 )^2=49152$ quasi-isotropic locations. At this resolution, the average distance between the center positions was almost an order of magnitude less than the average distance between the 542 GRBs. 

During the analysis, we assumed that the distribution was factorizable, {{i.e.,}} the sky exposure (the sensitivity of the experiment in detecting GRBs having measurable redshifts) was independent of~$z$ during the analysis. The statistical analysis by \citet{universe8070342} supports this assumption. Because of this, it was possible to evaluate variations in the spatial distribution using a spherical cap by assuming that GRBs outside the slice had the same sky exposure as those inside the slice. 

To generate random catalogs, we mixed the events' angular and radial locations, creating datasets in which the angular and radial distributions were independent. 
Using \mbox{2000 random} catalogs, we determined the $K$ value inside a sphere cap of a given size, $A$, and the redshift slice size $n$ and compared them with the greatest number of GRBs ($K$) with the \mbox{observed distribution}. 


As we pointed out above, mixing the angular and radial parameters of the data is the only way to create a statistically correct random catalog as the redshift observations' sky exposure map is not known, i.e., we do not know the probability of observing a GRB with a redshift in a given direction. It is well known that this kind of randomization will generate higher fluctuations than in the original data, but here we needed to use this sub-optimal method as the sky exposure map was not known. 

\section{The Northern Hemisphere GRB Distribution}
\label{sec:result}

The probability of measuring the value $K$ for a certain $n$ and $A$ was obtained by applying the bootstrap method described in the previous section. 
This was performed by retaining the sky position of each GRB and replacing its redshift with that of another GRB selected randomly from the sample. 
This was performed for all 262 GRBs in the northern galactic hemisphere.  
This process was repeated $W=2000$ times so that the number of times $K$ was greater than or equal to the measured value of $K$ could be obtained. 
This frequency estimated the probability of measuring a count of $K$ or larger.
This number was denoted as $P$, and $p=P/W$ estimated the probability. 
 
\begin{figure}[H]
    \includegraphics[width=0.9\columnwidth]{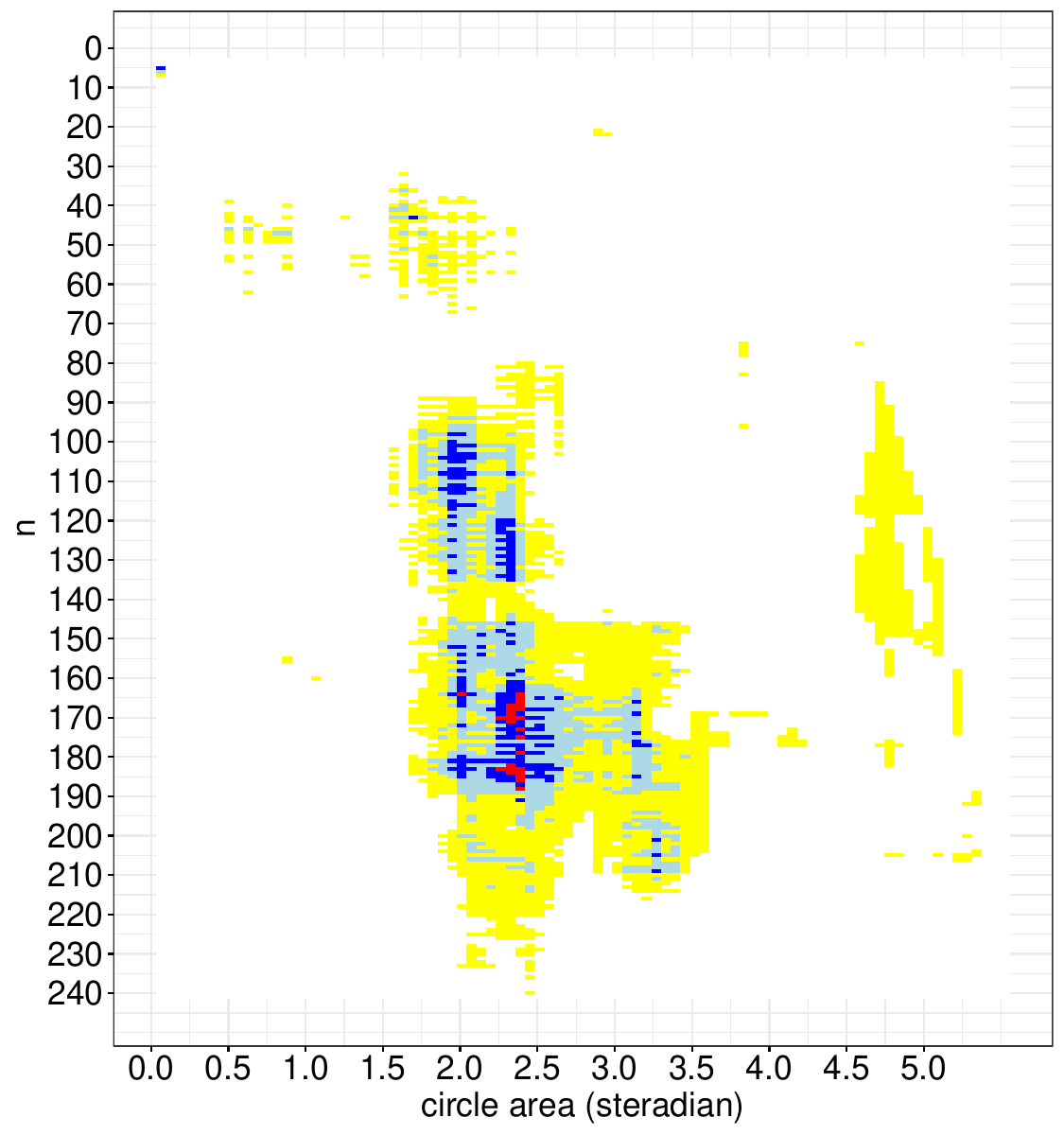}
    \caption{The bootstrap probabilities from the northern hemisphere on the (n, A) plane. White: $p \ge 0.2$; yellow: $0.2>p \ge 0.1$; light blue: $0.1>p \ge 0.05$; blue: $0.05>p \ge 0.02$; red: $0.02>p \ge 0.0$. Only blue and red blocks were significant.
    }
    \label{fig:EszakProb}
\end{figure}

Figure~\ref{fig:EszakProb} demonstrates the result of this analysis. 
There were four separate areas in the $n$, $A$ plane in which clustering was found to occur, as indicated by deviations from isotropy/homogeneity. Notice that only the dark blue and red dots were considered to be significant ($p<0.05$).

The first of these clusters occurred for $n=5$ and $A=0.0628$ ($0.02\pi$ steradians), with an observed $K=4$. Only one case ($5$, $0.0628$) reached the $p<0.05$ significance level in our analysis, with the bursts found within a redshift range of $0.59 \le z \le 0.62$. 
Four of the five GRBs in this group were located close together (further information can be found \mbox{in \citet{horv24MNRAS}). }

The second and third clusters in this parameter space occurred at around $41 \le n \le 47$ and $1.6 \le A \le 1.9$, with $K$ typically being $29 \le K \le 32$ for the second and $97\le n \le 113$ and $1.9 \le A \le 2.1$ for the third cluster. These clusters were found in the redshift range of $0.9 \le z \le 1.3$. The former area was only marginally significant ($p \le 0.04-0.08$), while the latter was more significant ($p < 0.05$). The sky area of these clusters corresponded to the position of the Hercules--Corona Borealis Great Wall \citep{hhb14,hbht15,2020MNRAS.498.2544H}. For an additional analysis of these three clusters, see \citet{horv24MNRAS}.

The fourth new cluster occurred at around $n=163-190$ and $A=2.0-2.5$.  
This cluster contained 105 pairs of $(n,A)$ for which the frequency was less than $5\%$. 
{Among these pairs, there were 72, 46, and 24 pairs of $(n,A)$, which had frequencies of less than $4\%$, $3\%$, and $2\%$, respectively.}
The smallest frequency was $0.75\%$ where $n=168$ and $A=2.39$, with $K=114$. These parameters indicate that in the redshift range of $z_{\rm start} \le z \le z_{\rm end}$ (corresponding to $n$-values of $n_{\rm start} \le n \le n_{end}$), there were 168 consecutive GRBs, where a spherical cap with an area of $2.39$ steradians contained 114 GRBs within it. In this special case, the closest GRB among the 168 was the 21st-closest GRB, and the farthest was the \mbox{189th-closest GRB.} 

\begin{figure}[H]
    \includegraphics[width=1.0\columnwidth]{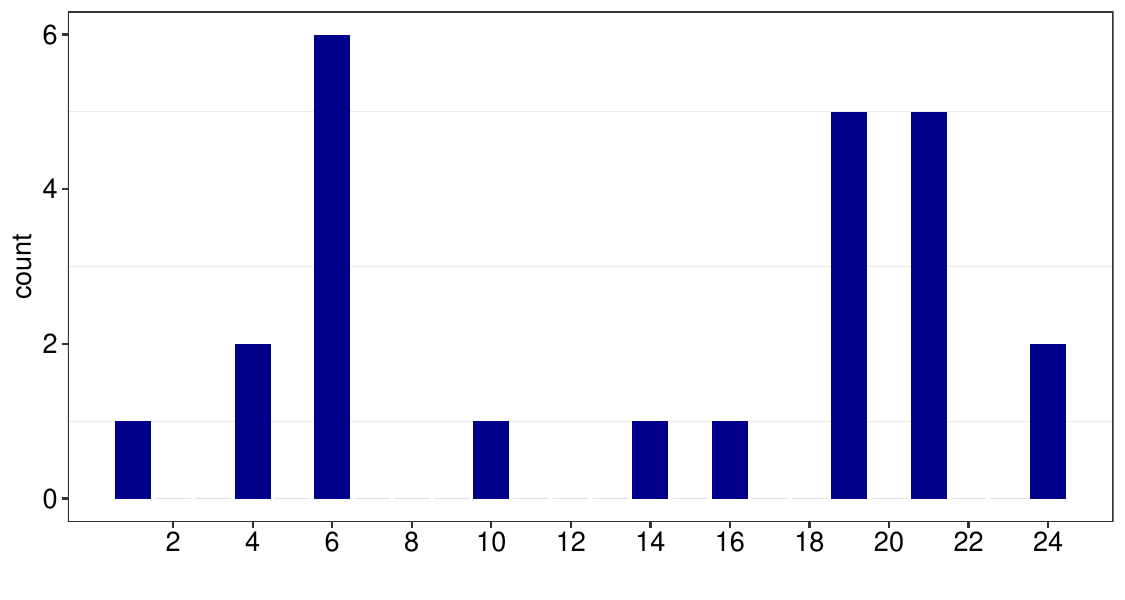}
    \includegraphics[width=1.0\columnwidth]{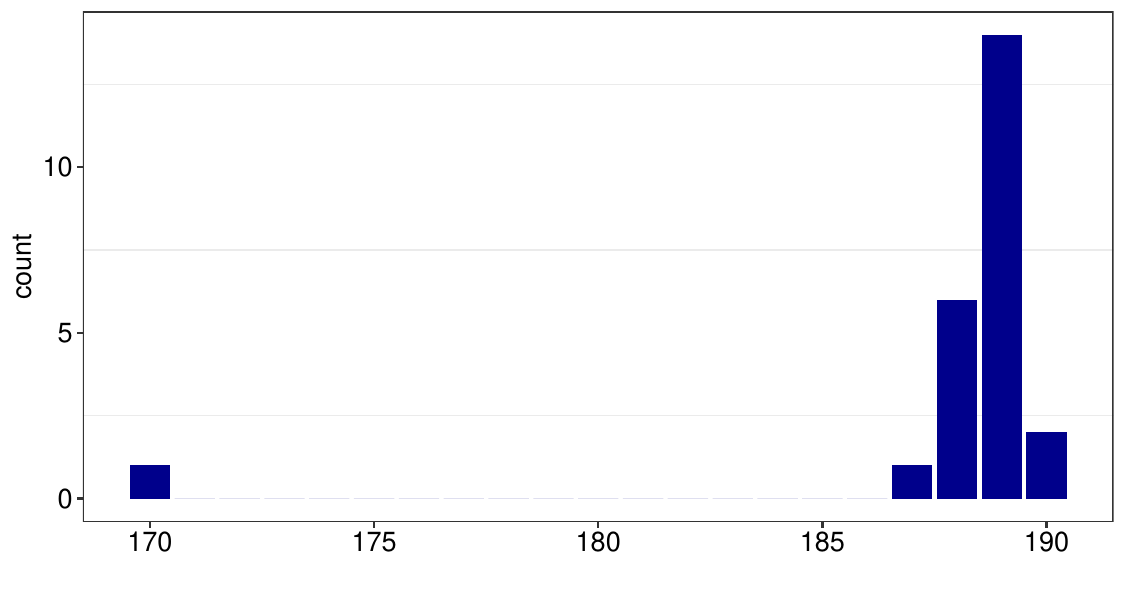}
    \caption{An example of an anisotropic distribution spanning 24 points in Figure \ref{fig:EszakProb}, where the probability is less than $0.02$. The top panel shows the distribution of the starting points, while the bottom panel shows the distribution of the ending points. See the text for more details.
    }
    \label{fig:kezdo24vegis}
\end{figure}

\begin{figure}[H]
    \includegraphics[width=0.9\columnwidth]{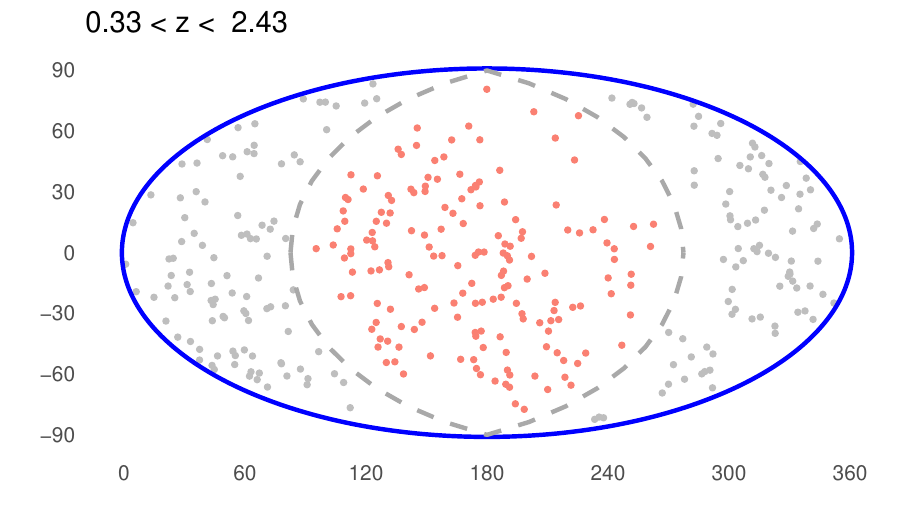}
    \caption{{An} 
    example of a large sky distribution anisotropy. The start and end points are $n_{\rm start}=19$ and $n_{\rm end}=189$, indicating that the anisotropy lies in the redshift range of $0.33 < z < 2.43$. {Red dots are the GRBs on the northern hemisphere, greys are on the southern.}    }
    \label{fig:egen033z243}
\end{figure}

Twenty-three other $(n,A)$ pairs had probabilities less than $0.02$. The distribution of these 24 radial slices' starting points can be seen in Figure~\ref{fig:kezdo24vegis}. This figure also shows the distribution of the 24 end points ($170 \le n_{\rm end} \le 190$). The start of the radial slices peaked
around ${n_{\rm start}=6}$, $n_{\rm start}=19$, and $n_{\rm start}=21$, while the end points were much more consistent, with more than half of the values being around $n_{\rm end}=189$.
An example of an anisotropic GRB distribution region is shown in Figure~\ref{fig:egen033z243}, occurring between the starting point of $n_{\rm start}=19$ ($z_{\rm start}=0.33$) and the end point of $n_{\rm end}=189$ ($z_{\rm end}=2.43$).

\section{Differences Between the North and South Celestial Hemispheres}

\citet{horv24MNRAS} compared the $K(n,A)$ distributions of the two hemispheres. \linebreak {Figure~13 in} 
that paper demonstrates that the two hemispheres exhibited similar, large-scale isotropy over most of the $(n, A)$ plane
where the differences were small. 
This means that the observed K distributions in the $(n, A)$ plane were similar in both hemispheres. 
However, there was a large region in the northern hemisphere in which the distributions differed by more than 14 GRBs with a certain radial slice and spherical cap.
This suggests that the northern hemisphere contains a large area in which there are significantly more GRBs than one would expect for a homogeneous, isotropic distribution. However, \citet{horv24MNRAS} did not explore the differences for 
$n>119$ and thus was not able to determine the upper limit on the cluster's extent in the existing GRB data. 

\begin{figure}[H]
    \includegraphics[width=\columnwidth]{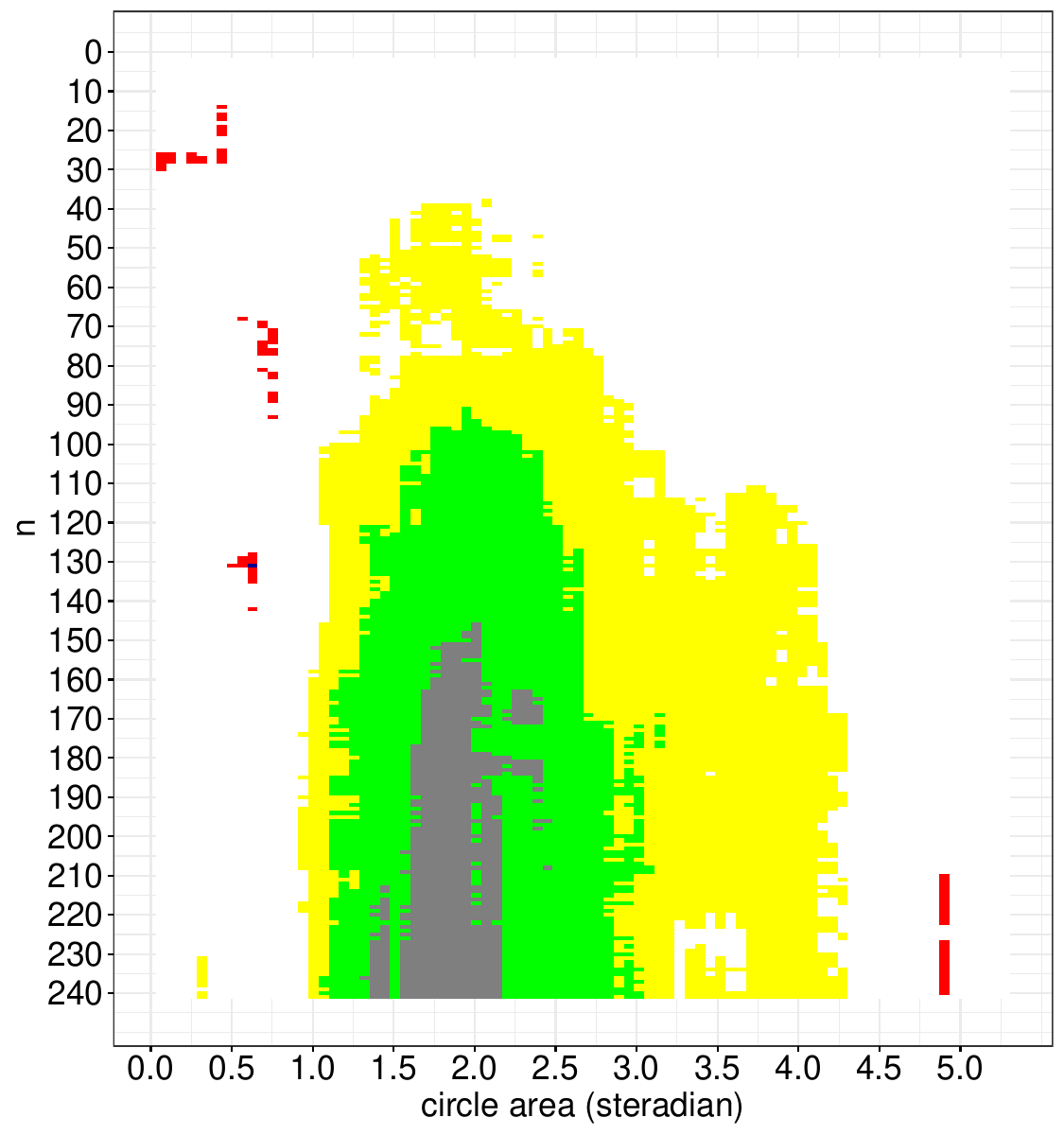}
    \caption{The difference between the $K_n$, northern, and $K_s$, southern, hemispheres' $K(n,A)$ functions. 
    Blue: $K_n-K_s = -3$; 
    red: $K_n-K_s = -2$;
    white: $-2 < K_n-K_s < 6$;
    yellow: $5 < K_n-K_s < 13$;
    green: $12 < K_n-K_s < 21$; and
    gray: $20 < K_n-K_s$.}
    \label{fig:KivonPboM}
\end{figure}

The extended analysis described in this paper allowed the upper limit on the size of this cluster to be explored in greater detail. Although the $z$ range was expanded, Figure~\ref{fig:KivonPboM} demonstrates that it was still difficult to identify an upper limit on the cluster size. 
This seems to be due, more than anything, to the fact that the dataset only contained 262 GRBs. When a cluster size of $n=242$ was used, there were only 20 possible starting points by which to study the redshift range.
However, the difference between the K-value in the northern galactic hemisphere, $K_n$, and the K-value in the southern galactic hemisphere, $K_s$, was still significant for a large $n$. 

\begin{figure}[H]
    \centering
    \includegraphics[width=\columnwidth]{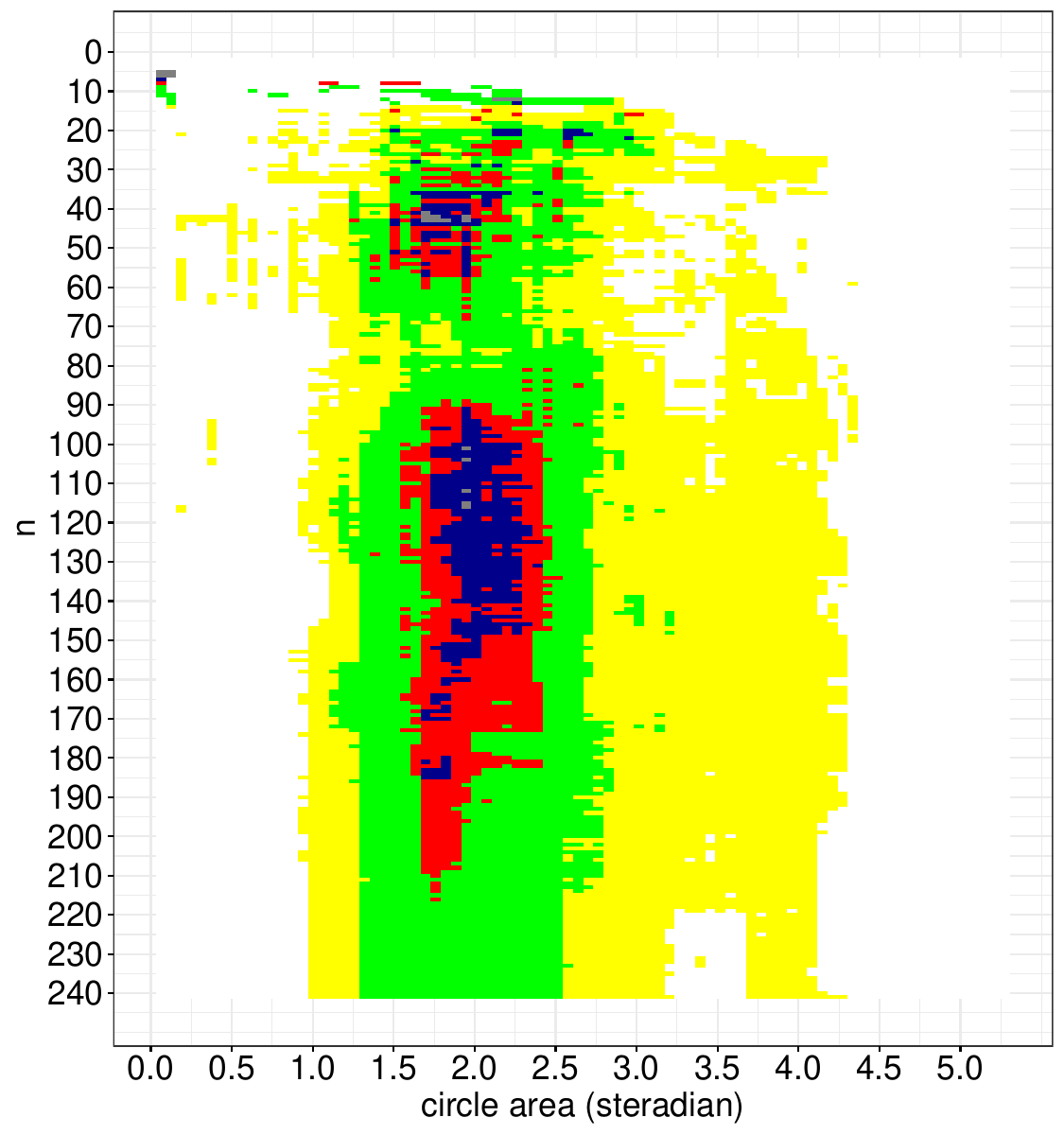}
    \caption{The relative difference between K measured in the northern and southern galactic hemispheres (${d = {(K_n - K_s)} / n}$).
  White: $ d < 0.03$;
   yellow: $0.03 < d < 0.075$;
    green: $0.075 < d < 0.115$; \linebreak 
    red: $0.115 < d < 0.135$;
     blue: $0.135 < d < 0.16$; and
    gray: $0.16 < d $.}
    \label{fig:RelPperM}
\end{figure}

We noted that $K_n$ and $K_s$ increased at a large $n$, as well as their difference.
We defined the relative difference $d$ between the K-values as $d = (K_n-K_s)/n$. The relative difference as a function of $n$ and $A$ is demonstrated \mbox{in Figure~\ref{fig:RelPperM}.}

Because the relative difference between the hemispheres decreased at a large $n$ ($n>140$), we inferred that the distribution approaches homogeneity on the scale of the full sky. This is also consistent with our observation that there were no obvious selection biases, such as undersampling a particular hemisphere due to sky exposure, that contributed to \mbox{large-scale anisotropies.}

\section{Discussion}

In this work, we continued to search for GRB clusters using the methods described by \citet{horv24MNRAS}. By concentrating the analysis on the northern galactic hemisphere, we were able to find evidence that the HerCrbGW has a larger radial range than previously suspected.

In addition to the three clusters we found in our previous work, here we identified a fourth cluster containing $110 < K < 120$  GRBs (with a $z$ slice of $163\le n\le 190$) and occupying an angular area in the sky of $2.0\le A \le 2.5$. This cluster was found in the same direction as the second and third clusters. The fourth cluster encompassed the third one---in the same way that the third cluster encompassed the second one---and it spanned an even larger $0.33 \le z \le 2.43$ redshift range than either of the previous two clusters. The co-mingling of these clusters suggests that the HerCrbGW  is  
significantly larger in radial size than was previously identified. 

By extending this analysis to a very large scale, we recognize that we potentially introduced an additional bias. The process described here probed for overly dense volumes within a certain angular region and redshift slice. The number of points found inside and outside the spherical cap $A$ depended non-linearly on the angular two-point correlation function. The size of the circle acted as a cutoff for the filtering, causing the non-linear behavior of the distribution of $K$ values. The actual relationship between the spatial distribution and the angular two-point correlation function became further complicated as the angular diameter distance changed the spatial (co-moving) geometry in the investigated cone. 
This could have introduced additional Poisson noise that would have made it difficult to directly compare the results of the bootstrap point radius test to those for the spatial two-point correlation function.

One may {put} the following question: could a matter concentration of the size discussed in our paper be possible at all in the observable universe? 
Solving this problem, \citet{Eingorn:2016v} found  a size of such a concentration of 3700 Mpc. It is worth noting that this value fits with cosmic structures found by \citet{clo12}, \citet{hhb14}, and \citet{BalazsRing2015}, and their sizes essentially exceed the previously reported scale of homogeneity of $\approx 370$ Mpc \citep{2010MNRAS.405.2009Y}.

{\citet{racz2017AIP} has investigated the distribution of matter in the Millennium-XXL and Horizon Run cosmological simulations using the MCMC Metropolis–Hastings algorithm, finding that a significantly larger sample (8--10 thousand objects) would be needed to accurately detect matter's distribution on the baryon acoustic oscillation (BAO) scale. The current measurements of 542 GRBs are not sufficient to reliably sample BAO-scale structures. The sizes of the large-scale wall and ring-like structures detected previously are significantly larger than those found in simulations, suggesting that sampling with GRBs is currently limited to the largest cosmic structures. Simulation results with larger sample sizes are significantly different from a completely random distribution. 
}

Collisions between galaxies are an important source of enhanced star formation activity \citep{2019MNRAS.485.1320M}. The frequency of collisions depends on the square of the number density of the objects participating in collisions. Let us suppose a $\delta$ first-order perturbation in the $\nu_0$ homogeneous spatial number density, $\nu = \nu_0(1 + \delta)$, with the frequency of collisions being proportional to $\nu^2 \approx \nu_0^2(1 + 2\delta)$ in this order. Consequently, the amplitude of the increase in the collision frequency will be higher with a factor of two than that of the $\delta$ number density enhancement. Therefore, it may occur that one finds anomalies in GRBs’ spatial distribution at some level of significance which are not seen in other cosmic objects.

Consequently, large-scale anomalies in the GRB spatial distribution can exist which are not necessarily seen in other cosmic objects. 
Further detailed observations are necessary to obtain a satisfactory solution to this problem.

\section{Summary}

The discovery of the Sloan Great Wall (with a size of $\sim$\,0.4~Gpc) using galaxy locations \citep{Gott05} created the ability for observers to identify large-scale universal structures via luminous objects. The association of luminous quasars with galaxies has allowed even larger structures to be identified, including the Huge Large Quasar Group \citep{clo12} and the Giant Quasar Arc \citep{2022MNRASLopez} (each with a size of 1.2~Gpc).
Thanks to their high luminosities, GRBs have made it possible to study giant walls and filaments.
However, the number and sizes of these walls challenge standard cosmological models, unless their detection can be attributed to statistical flukes ({e.g.,} \cite{Li2015,2020A&A...633L..10T}).
For this reason, it is important to revisit and continue exploring claims of large-scale universal structures identified via GRBs.

In this manuscript, we applied the clustering detection method described in Section \ref{sec:method} to the GRB distribution in the northern galactic hemisphere. We found evidence supporting the existence of the HerCrbGW \citep{hhb14}. However, we also found evidence that the HerCrbGW is larger than previously identified and that it appears to encompass several smaller clusters. We do not hypothesize what this large structure might represent, other than commenting that it does not appear to be the result of either a statistical fluctuation or a known sampling bias. 
We do note, however, that applying this method to 
our sample with N=262 points 
could introduce spurious Poisson noise. 
This additional source of noise is capable of subtly altering the interpretation of what constitutes a cluster.
\vspace{6pt}

\authorcontributions{Conceptualization, {I.H., Z.B., L.G.B., J.H., B.K., I.I.R., P.V., and S.P.}; 
data curation, {I.H., Z.B., L.G.B., J.H., B.K., I.I.R., P.V., and S.P.}; formal analysis, I.H. and Z.B.; investigation, all; methodology, {I.H., Z.B., L.G.B., J.H., B.K., I.I.R., P.V., and S.P.}; project administration, S.P. and I.H.; resources, {I.H., Z.B., L.G.B., J.H., B.K., I.I.R., P.V., and S.P.}; software, I.H. and Z.B.; supervision, {I.H., Z.B., L.G.B., J.H., B.K., I.I.R., P.V., and S.P.}; validation, {I.H., Z.B., L.G.B., J.H., B.K., I.I.R., P.V., and S.P.}; visualization, S.P., Z.B., L.B., and I.H.; writing---original draft, {I.H., Z.B., L.G.B., J.H., B.K., I.I.R., P.V., and S.P.}; writing---review and editing, {I.H., Z.B., L.G.B., J.H., B.K., I.I.R., P.V., and S.P.} All authors have read and agreed to the published version of the manuscript.}

\funding{This work was partially supported by the EKOP-24-4-II-48 University Research Scholarship Program and Project Nos. TKP2021-NVA-16 and TKP2021-NKTA-64, implemented with funding provided by the Ministry of Culture and Innovation of Hungary from the National Research, Development and Innovation Fund.}


\dataavailability{The data underlying this paper are available in the Gamma-Ray Burst Online Index (GRBOX) database published by the Caltech Astronomy Department (\url{http://site.astro.caltech.edu/grbox/grbox.php}),
Jochen Greiner's table (\url{https://www.mpe.mpg.de/~jcg/grbgen.html}), 
and the relevant Gamma-ray Coordination Network (\url{https://gcn.gsfc.nasa.gov/gcn3_archive.html}) messages. The code and data used for this study are available at \url{https://github.com/zbagoly/GRBSphericalCapStat}.}

\conflictsofinterest{The authors declare no conflicts of interest.} 
\printendnotes 
\begin{adjustwidth}{-\extralength}{0cm}

\reftitle{References}

\PublishersNote{}
\end{adjustwidth}
\end{document}